\documentstyle[aps,prl,multicol,fancyheadings,epsfig,graphicx,amsbsy,amssymb,amstex]{revtex}

\def\bbbc{{\mathchoice {\setbox0=\hbox{$\displaystyle\rm C$}\hbox{\hbox
to0pt{\kern0.4\wd0\vrule height0.9\ht0\hss}\box0}}
{\setbox0=\hbox{$\textstyle\rm C$}\hbox{\hbox
to0pt{\kern0.4\wd0\vrule height0.9\ht0\hss}\box0}}
{\setbox0=\hbox{$\scriptstyle\rm C$}\hbox{\hbox
to0pt{\kern0.4\wd0\vrule height0.9\ht0\hss}\box0}}
{\setbox0=\hbox{$\scriptscriptstyle\rm C$}\hbox{\hbox
to0pt{\kern0.4\wd0\vrule height0.9\ht0\hss}\box0}}}}
\newcommand{\beq}{\begin{eqnarray}} 
\newcommand{\eeq}{\end{eqnarray}} 
\newcommand{\tp}{{t^\prime}} 
\newcommand{\tap}{{\tau^\prime}} 
\newcommand{\oJ}{\omega_J} 
\newcommand{\Dp}{\Delta \phi} 
\newcommand{\nnn}{\nonumber\\} 
\newcommand{\bk}{{\bf k}}
\newcommand{\bp}{{\bf p}}

\begin{document}
\title{Probing Pseudogap by Josephson Tunneling}
\author{Ivar Martin and Alexander Balatsky}
\address{Theoretical Division, Los Alamos National Laboratory, 
	Los Alamos, NM 87545}

\date{Received \today }

\maketitle

\begin{abstract}
We propose here an experiment aimed to determine whether there are 
 superconducting pairing fluctuations in the pseudogap regime of the high-$T_c$ materials.
In the experimental setup, two samples above $T_c$ are brought into 
contact at a single point and the 
differential AC conductivity in the presence of a constant applied bias 
voltage between the samples, $V$, should be measured. 
We argue the the pairing fluctuations will produce randomly
 fluctuating Josephson current with 
zero  mean, however the current-current correlator  
 will have a characteristic frequency given
by Josephson frequency $\omega_J = 2 e V /\hbar$.  We predict that   the 
differential AC conductivity should have a peak at the Josephson 
frequency   with the width determined 
by the phase fluctuations time.
\end{abstract}
\pacs{Pacs Numbers: XXXXXXXXX}

\vspace*{-0.4cm}
\begin{multicols}{2}

\columnseprule 0pt

\narrowtext
\vspace*{-0.5cm}

One of the long-standing puzzles of the high-temperature superconductivity 
is the nature of the so-called pseudogap regime.  The pseudogap 
regime occurs in the wide range of temperatures above superconducting transition
temperature in the underdoped cuprate superconductors.  
It is characterized by the suppressed quasiparticle density of 
states in the vicinity of the Fermi level. The similarity of the density of states 
in the pseudogap regime and in the superconducting state 
lead many \cite{kivelson} to believe that that the pseudogap 
itself is of a superconducting origin.  In this view, the long range superconducting
order in the pseudogap regime is destroyed by phase fluctuations. However, locally,
both electronic pairing and fluctuating regions of superconductivity should 
persist.  
Therefore, in this picture, the superconducting phase transition is
 believed to be a superconducting phase-ordering transition.  Moreover, recent experiments by Orenstein
  and collaborators claim that local superfluid density is present even above 
  $T_c$ in Bi2212 materials \cite{Orenstein}.

Whether or not the pseudogap indeed has a superconducting origin remains to be 
verified experimentally.  Such standard experimental test as the vanishing 
resistivity,
or the Meissner effect are bound to fail.  Both these test require spatial and
temporal stability of the superconducting phase on the time scale of the experiment.  
On the other hand, the pseudogap state can at best have superconducting order 
parameter that varies both in space and in time. 
A successful test may be 
possible if the fluctuations could somehow be stabilized by the proximity to a ``real'' 
superconductor \cite{janko}.  Another approach is to probe superconductivity locally 
in space on the time scales comparable or shorter than the characteristic 
time of the phase fluctuations. 

In this letter we propose the first experiment of 
this type, which is based on the AC Josephson effect. 
The main point we make  is that in the presence of phase fluctuations 
Josephson current $j(t)$ across the tunneling contact is a random time-dependent quantity. It has a dispersion
that is the current-current correlator $\langle j(t)j(t') \rangle$,
 which is related to the corrections to the conductivity across the junction.
  This  current-current correlator ``remembers'' about its Josephson origin and has a scale set by Josephson frequency
   and phase fluctuation time. Therefore, the conductivity of a junction 
will have a correction
   due to current fluctuations,
   \beq\label{main1}
   \Delta \sigma \propto   1/V ,   
   \eeq
 where $V$ is the applied voltage across the junction.   
 The crucial new aspect of the proposed 
approach is that we will focus on the characteristic time scale
 of frequency fluctuations, assuming that tunneling occurs
in a small region where the spatial dependence can be ignored. 
We will focus on the time dynamics of the   phase fluctuations 
in our analysis.

\begin{figure}[htbp]
  \begin{center}
    \includegraphics[width = 3.0 in]{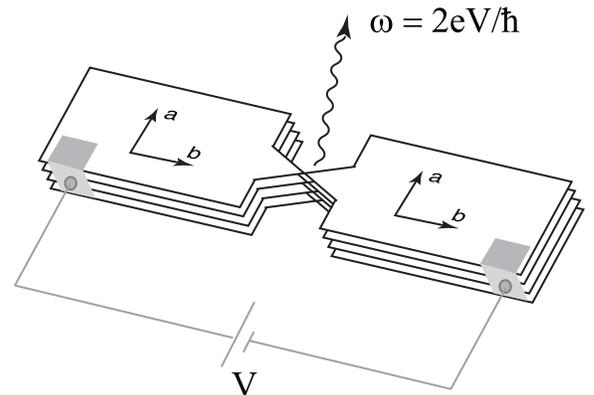}
    \caption{Consider two parts of a superconductor brought into a tunneling contact
at a temperature $T>T_c$, so that the phase of the order parameter is
no longer stationary.  Our approach is to consider the Josephson current (Eq. 
\protect{\ref{j1}}) as a randomly fluctuating quantity, with the three major sources of 
randomness: the amplitude noise, the frequency noise, and the phase noise.  
We also assume that  the contact is small enough for the superconducting 
parameters to be spatially uniform in the vicinity of the contact.}
    \label{fig1}
  \end{center}
\end{figure}

When two  pieces of a superconductor are joined by a weak link, a superconducting 
current begins to flow through the link in the absence of applied voltage 
between the superconductors.  The current is related to the difference of the
phases $\phi_1$  and $\phi_2$ of the superconductors,
\beq\label{j1}
j = j_0 \sin(\phi_1 - \phi_2),
\eeq
with the parameter $j_0$ related to the coupling strength between the two superconductors.  
For a superconductor in an equilibrium, evolution of the phase,
\beq
{\partial\phi}/{\partial t} = -2\mu/\hbar,
\eeq 
is determined\cite{anderson} by the superconductor chemical potential, $\mu$.
Therefore, the phase as a function of time is 
\mbox{$\phi(t) = -2\mu t/\hbar + \phi_0$}, 
where $\phi_0$ is the phase at time $t = 0$.  For two coupled superconductors 
the phase difference evolves as
\beq
\phi_1(t) - \phi_2(t) = 2(\mu_2 - \mu_1) t /\hbar  + \phi_1(0) - \phi_2(0).
\eeq
The difference of the chemical potentials equals the applied voltage, 
$\mu_2 - \mu_1 = e V$.  Hence, if $V = 0$, both the phase difference and the current 
given by Eq.~(\ref{j1}) remain constant.  However, in the presence of a bias, $V$, 
the phase difference  grows linearly with time and the current oscillates according to 
\beq
j(t)  = j_0 (\omega_J) \sin(\omega_J t + \Delta \phi),
\eeq
with  the frequency $\omega_J = 2 e V/\hbar$ and the initial phase 
$\Delta \phi = \phi_1(0) - \phi_2(0)$.  The effect of generating an alternating 
current by applying a constant bias to a superconducting tunnel junction is called 
AC Josephson effect.  An important feature of this effect is that the frequency of
the generated current is only a function of the applied bias voltage and is 
independent of the microscopic and macroscopic parameters of the system.  The scale
of the frequency is about 0.5 $\Gamma$Hz per 1 microvolt.
The AC Josephson effect is routinely observed in the superconducting regime.  It can
be observed either directly as the micro-wave emission from the oscillating current, or 
indirectly as ``Shapiro steps'' \cite{shapiro} in the DC $I-V$ curves measured 
in the presence of the oscillating bias voltage component.   

For AC Josephson effect
to be observable in the pseudogap regime, the measurement has to be local both in space 
and time. 
Suppose that above the transition temperature, superconductor can be modeled as a 
collection of superconducting islands of a characteristic size $L$,  inside 
which phase fluctuates at a rate $\Lambda$.  Then, if the size of a contact between two 
superconductors is less than $L$ then the superconducting state is essentially uniform in the
vicinity of the contact.  If the applied bias voltage is such that  the 
Josephson frequency is larger than $\Lambda$ then the Josephson oscillations are 
faster than the phase fluctuations dynamics, and hence can be approximately modeled as
\beq\label{j2}
j(t) = j^*(\omega_J, \Lambda, L) \sin(\omega_J t + \Delta \phi (t)),
\eeq
with the amplitude $j^*$ being renormalized by spatial and temporal fluctuations of the 
superconducting phase.  In general, the Josephson frequency can also fluctuate around
its average value due to voltage fluctuations which are particularly important for small
samples.  However, we assume that these effects can be absorbed into the overall fluctuations
of the phase, $\Delta \phi (t)$.
It is important to note that both $\Lambda$ and $L$ are functions of 
temperature, with L and $1/\Lambda$ diverging at the superconducting transition.
The parameter $L$ is related to the phase gradient correlation function $W$ 
considered by Franz and Millis \cite{millis}.  Relationship between $\Lambda$ and $L$ is 
a subject of an active interest.  In the vortex diffusion picture, where phase fluctuations
are produced by moving vortices,  $L$ is a distance 
vortex travels during the time $1/\Lambda$, namely $L = \sqrt{D/\Lambda}$.  Here $D$
is the vortex diffusion constant.  This corresponding dynamical critical exponent is 
$z = 2$.  Alternatively, if the phase fluctuations are governed by fast ballistic 
dynamics, the relation between the parameters $L$ and $\Lambda$ should be $L \Lambda = v^*$,
where $v^*$ is propagation speed for the ballistic modes. This corresponds to $z = 1$.
Using different geometries in the 
experimental setup that we propose below may help to determine the relevant model.

A possible experimental setup that can be used to perform the measurement of the 
AC Josephson effect in the pseudogap regime is shown in the figure \ref{fig1}.  
The crucial aspect is that the point of contact between the superconductors be as small as 
possible.  If the size of the contact becomes larger than $L$ then
in addition to the temporal phase fluctuations a the point of contact one needs to include 
spatial fluctuations, which can lead to a significant suppression of the effect.
Another desirable feature is that the two superconductors be only a few  $ab$-planes thick.
This is because the size of the superconducting ``islands'' is  likely to be more extended 
in the planes, compared to across the planes.  Hence, we believe that the effect
we propose is more likely to be observed in the geometry of figure \ref{fig1}, 
although c-axis tunneling may also yield similar results.  Finally, using very thin samples
reduces the transition temperature \cite{1-layerTc}, thereby making the pseudogap regime 
accessible at lower temperatures, where the thermal fluctuations are reduced.

There are several ways the  oscillating super-current in the pseudogap regime 
can be detected.  Here we consider two methods: 1) differential AC conductivity measurements
in the presence of constant bias voltage, 2) detection of electro-magnetic radiation 
generated by the oscillating Josephson current.  Although there is no coherent Josephson 
current in the pseudogap regime, the junction is expected to have strong response 
to the perturbations acting at the frequencies near $\omega_J$.  Such super-current is also 
expected to generate a radiation  peak at the frequency $\omega_J$, with a width of the peak 
governed by the phase fluctuations.

\begin{figure}[htbp]
  \begin{center}
    \includegraphics[width = 3.0 in]{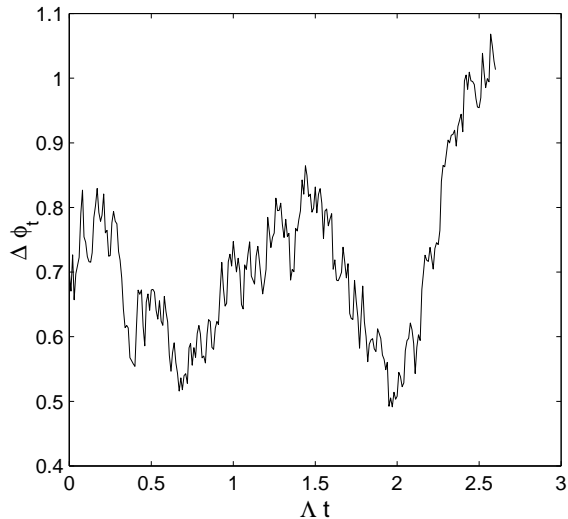}
    \caption{We assume here that the phase difference, $\Dp_t$,  between the superconductors forming the 
Josephson junction follows a one-dimensional geometrical Brownian walk as a function of time, $t$.
Such process is defined by a linearly increasing with time dispersion, Eq. (\protect{\ref{eq:diff}}).
  The initial phase, $\Dp_0$, is
also a random quantity evenly distributed in the interval $[0, 2\pi]$.}
    \label{fig:diff}
  \end{center}
\end{figure}

To make our qualitative arguments more formal we have to assign a particular form to the 
phase fluctuations.  Here we make an  assumption that the the phase difference 
between the superconductors, $\Delta \phi (t)$ follows a diffusion
process, as shown in Fig.~\ref{fig:diff}, with a variance 
\beq\label{eq:diff}
\langle (\Delta \phi_t - \Delta \phi_{t^\prime} )^2\rangle = 2 \Lambda |t - t^\prime|.
\eeq
and the  initial phase $\Delta \phi_0$ distributed uniformly in the interval 
$[0, 2\pi]$.  
The factor of 2 appears because for a weak tunnel junction the phases of on the both 
sides of the junction fluctuate independently, each at a rate $\Lambda$.

Viewing the Josephson current of Eq. (\ref{j2}) as a random quantity with a zero mean,
 we  can characterize it by its dispersion and autocorrelation.  The autocorrelation, 
according to the Kubo formula, determines the correction to the conductivity 
due to the fluctuating Josephson tunneling,  
\beq
\Delta \sigma (\omega) =  \frac {1} {\omega\nu}\int^t _{-\infty} {e ^{i \omega(t - \tp)}
	\langle j (t) j(t^\prime)\rangle d\tp},
\eeq
% Mahan 3.8.8
where brackets correspond to the $\phi$-averaging, and averaging over time $t$ is implied. 
Volume $\nu$ is necessary for normalization.
Substituting  expression for the current from Eq. (\ref{j2}), we obtain
\beq
\Delta \sigma &=& \frac {j^{*2}} {2\omega\nu} \int^t _{-\infty} 
	{e ^{i \omega(t-\tp)}
	\langle  \cos(\oJ (t- \tp) + \Dp_t - \Dp_\tp) -}\nnn
	&& \cos(\oJ (t + \tp) + \Dp_t + \Dp_\tp) \rangle d\tp. 
\eeq
Since $\Dp_t + \Dp_\tp = (\Dp_t - \Dp_0) + (\Dp_\tp - \Dp_0) + 2\Dp_0$, after averaging
over $\Dp_0$ in the interval $[0, 2\pi]$,
the second cosine disappears.  To average over 
$(\Dp_t - \Dp_\tp)$,  we invoke the relation 
$\langle \exp(i u)\rangle = \exp(-\langle u ^2\rangle/2)$,  valid for any 
normally distributed variable  $u$  with a mean zero.  Then after the integration we obtain
\beq
\Delta \sigma   =  \frac {j^{*2}} {4  \omega\nu} \left[
\frac{1}{\Lambda + i\omega - i\oJ} + \frac {1}{\Lambda + i\omega + i\oJ}
\right].
\eeq
As expected, the the real part of the conductivity, 
\beq\label{Res}
{\rm Re}(\Delta \sigma) = \frac {j^{*2} \Lambda} {4  \omega\nu}
\left[ \frac{1}{(\omega - \oJ)^2 + \Lambda^2} +  
	\frac{1}{(\omega + \oJ)^2 + \Lambda^2}\right],
\eeq
which corresponds to the in-phase response, has two peaks near $\pm\omega_J$.  The divergence 
as $\omega \rightarrow 0$ has no physical meaning, since no superconductivity related 
response is expected on the time scales larger than the characteristic phase fluctuation
time.  This translates into the condition $\omega \gtrsim \oJ$ for validity of Eq. (\ref{Res}).  
The imaginary part of the conductivity in the vicinity of $\oJ$ is about two times smaller
than the real part, and hence can be neglected in the total conductivity 
\mbox{${\rm Abs}\ \sigma= \sqrt{({\rm Re}\ \sigma)^2 + ({\rm Im}\ \sigma)^2 }$}.

Therefore, we predict that if the pseudogap regime is superconducting in origin there should be a peak in 
the differential AC conductivity at the frequency $\oJ \propto  V$, with the peak value
that scales as shown in Eq. (\ref{main1}): 
\beq\label{main}
\Delta \sigma \propto j^{*2}/\Lambda\oJ \propto 1/V .
\eeq
This is the main result of this paper.
While this correction may be small relative to
the normal (single-electron) current component, it can be extracted from the background conductivity
due to its extremely high sensitivity an applied external magnetic field.
As is evident, the magnitude of the correction is inversely proportional to the phase-breaking rate, 
$\Lambda$, and as a consequence should be more easily observable at temperatures close to the superconducting 
transition.  Consequently, a possible experimental approach is to start arbitrarily close to $T_c$ 
and to measure the microwave radiation from the weakly dephased Josephson current, and or to measure the 
differential AC conductivity as proposed above.  Then, gradually incrementing the temperature one can probe 
how the spatial and temporal fluctuations of the order parameter phase grow with the temperature.

In fact, a similar fluctuational AC Josephson effect can be searched for even in 
the conventional, ``low-$T_c$,'' 
superconductors, in the so-called paraconductivity regime \cite{AL}.  The paraconductivity regime
is characterized by superconducting order parameter fluctuations above $T_c$, and experimentally 
is associated with the rapidly decreasing (but finite) resistivity in the vicinity of $T_c$.  
Using the experimental setup we propose here, one could attempt to study the dynamics of 
the dephasing timescales in the close proximity of $T_c$ in the paraconductivity regime.
The difference between the conventional paraconductivity effect and the pseudogap is that the pseudogap 
is believed to extend far beyond the paraconductivity range where the rapid changes in the material 
resistivity occur.

Let us examine now more closely the assumptions that lead to the expression for the 
Josephson current, Eq.~(\ref{j2}).  Within the standard theory \cite{mahan}, the 
Josephson current is 
\beq\label{jFF}
j(t) = 2 e {\rm Im} [e^{-i \oJ t } \Phi_{\rm ret}(eV)],
\eeq
where the retarded correlation function $\Phi_{\rm ret}(eV)$ can be obtained from the 
Matsubara correlation function 
\beq
\Phi(i\omega)  = 2 \sum_{\bk\bp} T_{\bk, \bp} T_{-\bk, -\bp} \int_{\tau - \beta}^\tau {d\tap 
	F^\dagger(\bk, \tau, \tap) F(\bp, \tap, \tau)},\nonumber
\eeq
via analytical continuation $i \omega \rightarrow eV + i0$.  Here $T_{\bk, \bp}$ is a matrix 
element for tunneling from a state $\bk$ on one side of the
junction into a state $\bp$ on the other side of the junction, and 
$F(\bk, \tau, \tap) = \langle T_\tau c_{\bk\uparrow (\tau)} c_{-\bk\downarrow}(\tap) \rangle$
is an anomalous time-ordered Green functions.  
In Eq. (\ref{jFF}), we do not include the regular single electron contribution, proportional 
to $G(\bk, \tau, \tap) G(\bp, \tap, \tau)$.  The reason is that it does not carry the 
superconducting phase information and, therefore, does not produce the resonant features away from zero 
frequency.
In the absence of phase fluctuations $F$ is 
only a function of the time difference, 
$F^0(\bk, \tau-\tap) = (n_F(-E_\bk) e^{-E_{\bk}|\tau-\tap|} - 
n_F(E_\bk) e^{E_{\bk}|\tau - \tap|})/2E_\bk$.
Phase fluctuations can be incorporated phenomenologically into the anomalous Green functions as phase
factors
\beq
F(\bk, \tau, \tap) \rightarrow F(\bk, \tau, \tap) e^{i\phi(\tau, \tap)}.
\eeq
The form of the function $\phi(\tau, \tap)$ depends on the  
model of the phase fluctuations.  Here we assume that 
\beq\label{phi12}
\phi(\tau, \tap) = \phi(\tau) + \phi^\prime(\tau - \tap),
\eeq
where $\phi(\tau)$ and $\phi^\prime(\tau - \tap)$ are uncorrelated Brownian motions.  
Similar statistical
properties of $F(\bk,\tau, \tap)$ can be obtained \cite{dirk} from a gauge transformation of the
electron operators, $c_t = \psi_t e^{i \Theta(t)}$, under which 
$F(\tau, \tap) = \langle T_\tau \psi_{\bk\uparrow (\tau)} \psi_{-\bk\downarrow}(\tap) \rangle 
e^{i \Theta(\tau) + i \Theta(\tap)}$.  Since $\Theta(\tau) + \Theta(\tap)= 2 \Theta(\tau) - 
(\Theta(\tau)-\Theta(\tap))$, and under realistic assumptions 
($\Theta(\tau) - \Theta(\tap)$) is only a function of 
$(\tau -\tap)$, this approach yields an expression equivalent to Eq.~(\ref{phi12}), except for the 
correlations induced between the functions $\phi(\tau)$ and $\phi^\prime(\tau - \tap)$. 
In what follows we assume for simplicity that the correlations are absent.  Then doing the average over
the Brownian random process $\phi^{\prime}$ and integrating over $\tap$, for the Josephson current 
we obtain
\beq
j(t) = {\rm Abs}[j_0(\oJ + i \Lambda)] sin (\oJ t + \Delta \phi(t)),
\eeq
which is identical to the form of the Josephson current conjectured in Eq. (\ref{j2}).
The function $j_0(z)$ is the analytical continuation of the function $j_0(\oJ)$ which determines 
the amplitude of the Josephson current in the absence of the phase fluctuations.  In the case of 
s-wave superconductivity with a constant gap $\Delta$, this function is 
$j_0(eV) = (\sigma_0\Delta/e) K(eV/2\Delta)$, defined in terms of complete
 elliptic integral $K(x)$. In the case of d-wave superconductor
 $j_0(eV)$ is also a  nontrivial function of the relative orientation between
 lattices in two crystals. Its specific form is not important for our
 discussion.
  Finally, we should 
mention that in the current-current correlator, both $\phi$ and $\phi^\prime$ averages should be done on 
the product of currents, while we have done the averaging over $\phi^\prime$ independently in $j(t)$ and
$j(t^\prime)$.  The qualitative results for conductivity, however, remain the same with the two peaks 
at the frequencies $\pm \oJ$.

In conclusion, we propose to test the relevance  of the  phase fluctuations
 scenario in the
 pseudogap regime of the high-$T_c$ superconductors by investigating fluctuating 
 Josephson current at $T>T_c$. We focus on the temporal fluctuations of the phase
 assuming small-contact tunneling to ignore spatial dependence of the phase. We argue
  that although phase fluctuations will
  yield zero mean Josephson current, its autocorrelation function will produce
   finite correction to the conductivity of {\em normal} current
    across the junction. AC conductivity will exhibit the peak at Josephson
     frequency $\omega_J = 2eV/\hbar$ with the width determined by the characteristic 
     phase fluctuation rate $\Lambda$. possible experimental test  could be to 
     measure the junction AC conductivity $\sigma(\omega)$ 
     in the presence of constant bias $V$ and determine if it has a peak at $\omega_J$. 
We predict specific   dependence $\delta \sigma(\omega_J) \propto V^{-1}$  of the peak.
 Specific experimental  set up is shown on Fig. 1.

We would like to thank D. Morr, M. Graff, L. Bulayevski, J. Eckstein, and  M. Maley for useful 
discussions.  This work was supported by the DOE.

%\vspace*{-2.2cm}

\end{multicols}

\end{document}